\documentclass[sigconf]{acmart}

\usepackage[utf8]{inputenc}
\usepackage{booktabs,multirow}

\usepackage{dsfont}
\usepackage{bbm}
\usepackage{arydshln}
\usepackage{hyperref}

\hypersetup{
    colorlinks=true,
    linkcolor=blue,
    filecolor=magenta,      
    urlcolor=cyan,
    }

\usepackage[normalem]{ulem}
\usepackage{xcolor}
\usepackage{color}
\usepackage{rotating}
\usepackage{amsmath}
\usepackage{array}
\usepackage{enumitem}
\usepackage[linesnumbered,algoruled,boxed,lined]{algorithm2e}
\usepackage{kotex}
\usepackage{balance}
\usepackage[short]{optidef}
\usepackage{graphicx}
\usepackage{tcolorbox}
\usepackage{hyperref}

\newcolumntype{P}[1]{>{\centering\arraybackslash}p{#1}}
\newcolumntype{R}[1]{>{\RaggedLeft\arraybackslash}p{#1}}

\newcommand{\ie}{{\it i.e.}}
\newcommand{\eg}{{\it e.g.}}

\newcommand{\ul}{\underline}{}

\copyrightyear{2025}
\acmYear{2025}
\setcopyright{cc}
\setcctype{by-nc-nd}
\acmConference[SIGIR '25]{Proceedings of the 48th International ACM SIGIR Conference on Research and Development in Information Retrieval}{July 13--18, 2025}{Padua, Italy}
\acmBooktitle{Proceedings of the 48th International ACM SIGIR Conference on Research and Development in Information Retrieval (SIGIR '25), July 13--18, 2025, Padua, Italy}\acmDOI{10.1145/3726302.3729948}
\acmISBN{979-8-4007-1592-1/2025/07}

\begin{document}

\title{DIFF: Dual Side-Information Filtering and Fusion for \\ Sequential Recommendation}

\author{Hye-young Kim}
\affiliation{
  \institution{Sungkyunkwan University}
  \city{Suwon}
  \country{Republic of Korea}}
\email{khyaa3966@skku.edu}

\author{Minjin Choi}
\affiliation{
  \institution{Samsung Research}
  \city{Seoul}
  \country{Republic of Korea}}
\email{min_jin.choi@samsung.com}

\author{Sunkyung Lee}
\affiliation{
  \institution{Sungkyunkwan University}
  \city{Suwon}
  \country{Republic of Korea}}
\email{sk1027@skku.edu}

\author{Ilwoong Baek}
\affiliation{
  \institution{Sungkyunkwan University}
  \city{Suwon}
  \country{Republic of Korea}}
\email{alltun100@skku.edu}

\author{Jongwuk Lee}\authornote{Corresponding author}
\affiliation{
  \institution{Sungkyunkwan University}
  \city{Suwon}
  \country{Republic of Korea}}
\email{jongwuklee@skku.edu}




\begin{abstract}
Side-information Integrated Sequential Recommendation (SISR) benefits from auxiliary item information to infer hidden user preferences, which is particularly effective for sparse interactions and cold-start scenarios. However, existing studies face two main challenges. (i) They fail to remove noisy signals in item sequence and (ii) they underutilize the potential of side-information integration. To tackle these issues, we propose a novel SISR model, \emph{\textbf{D}ual Side-\textbf{I}nformation \textbf{F}iltering and \textbf{F}usion (\textbf{DIFF})}, which employs \emph{frequency-based noise filtering} and \emph{dual multi-sequence fusion}. Specifically, we convert the item sequence to the frequency domain to filter out noisy short-term fluctuations in user interests. We then combine early and intermediate fusion to capture diverse relationships across item IDs and attributes. Thanks to our innovative filtering and fusion strategy, DIFF is more robust in learning subtle and complex item correlations in the sequence. DIFF outperforms state-of-the-art SISR models, achieving improvements of up to 14.1\% and 12.5\% in Recall@20 and NDCG@20 across four benchmark datasets.
\end{abstract}

\maketitle
\section{Introduction}\label{sec:introduction}

Sequential Recommendation (SR)~\cite{FangZSG20SeqSurvey, WangHWCSO19SeqSurvey} aims to predict the next item the user will likely interact with by analyzing past user behavior. It is crucial in various web applications, including e-commerce and streaming services. Existing SR models employ diverse neural architectures to encode an item sequence into user representation. Among them, attention-based models~\cite{KangM18SASRec, SunLWPLOJ19BERT4Rec, kdd/MaKL19HGN} have shown outstanding performance gains by capturing intricate item correlations. However, these models only focus on item IDs, neglecting to utilize valuable item attributes.

Recently, \emph{Side-information Integrated Sequential Recommendation (SISR)}~\cite{Chang21NOVA, YuanDTSZ21ICAISR, ZhouWZZWZWW20S3Rec} addresses these limitations by modeling the item sequence using side-information. It incorporates various item attributes, \eg, ``Brand'' and ``Category'', into the recommendation process. SISR models demonstrate enhanced capability in capturing diverse collaborative signals across items, proving particularly effective in sparse user interaction and cold-start item settings.

Depending on item attribute fusion strategies, existing SISR models can be broadly categorized into three pillars: \emph{early}, \emph{late}, and \emph{intermediate fusion}\footnote{Although existing work~\cite{www/Wang24ASIF} designs it as hybrid fusion, it does not explicitly combine different fusion types. Thus, we rename it intermediate fusion to avoid ambiguity.}. Early fusion combines item ID and attribute embeddings before feeding to the model, enabling rich interactions across attributes. However, due to inherent differences in representation spaces, this simple aggregation may result in \emph{information invasion}~\cite{Chang21NOVA, XieZK22DIFSR}, in which the fused item characteristics become dominated or distorted by the ID or attribute information. Meanwhile, Late fusion~\cite{ZhangZLSXWLZ19FDSA} encodes IDs and attributes separately, delaying the fusion until the final prediction layer. Although each sequence is modeled effectively, it struggles to capture the correlation between item IDs and attributes. As an alternative, intermediate fusion~\cite{Chang21NOVA, XieZK22DIFSR, www/Wang24ASIF} computes the attention weight of attributes and leverages them to only guide the item correlation, preventing unnecessary interference between IDs and attributes.

\begin{figure}
\centering

\includegraphics[width=1.0\linewidth]{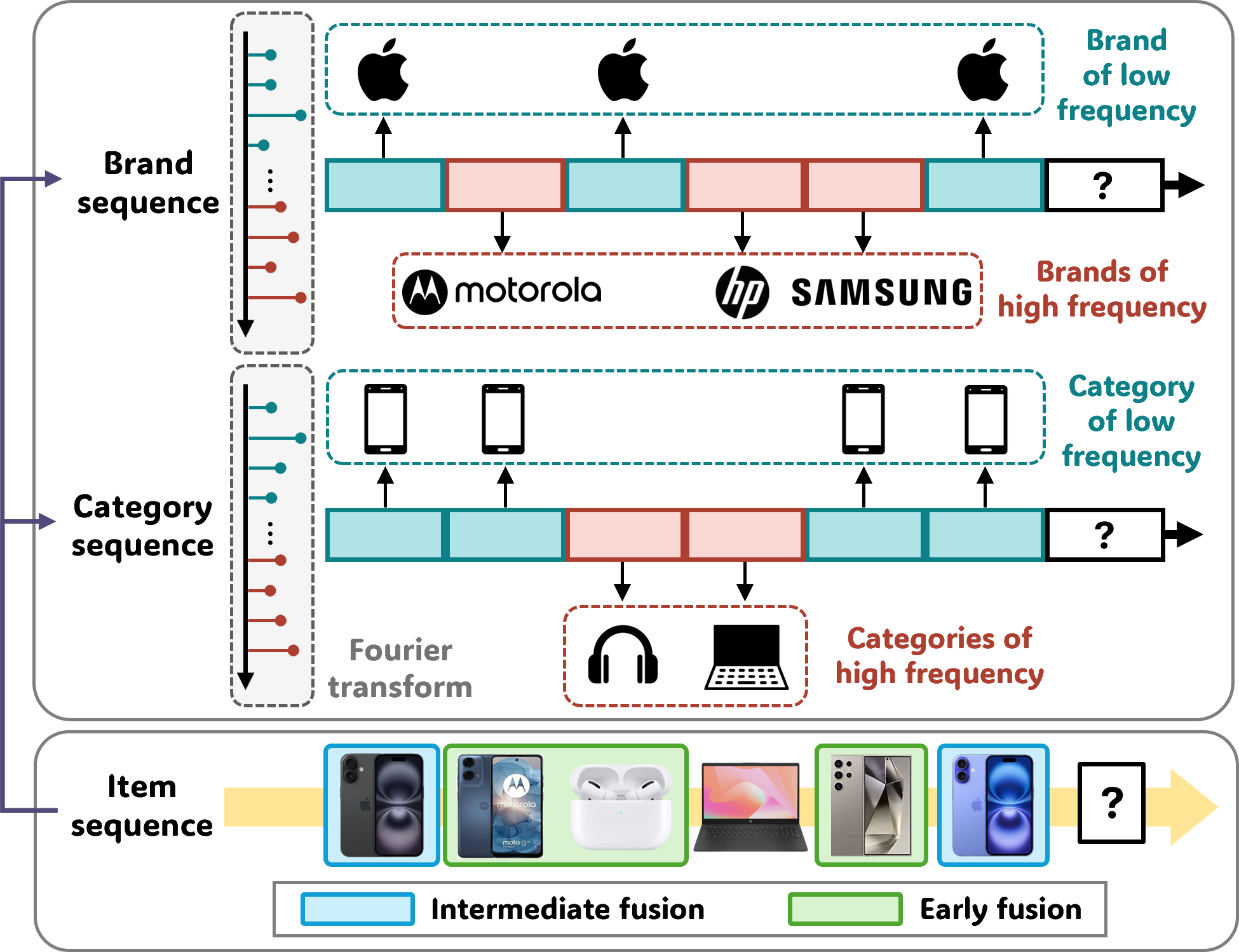}
\vspace{-5mm}
\caption{(i) Frequency signals and (ii) fusion types in side information integrated sequential recommendation. Frequency-based noise filtering removes the fourth item with inconsistent signals. Intermediate fusion (blue) highlights items aligned with key signals, while early fusion (green) captures broader combinations.}\label{fig:motivation}

\vspace{-3mm}
\end{figure}

While these fusion strategies have shown promising results in leveraging item attributes, they still face two critical challenges that need to be addressed.

\noindent
\textbf{(i) Noisy signals in item sequences}.
Item sequences often contain inconsistent patterns not aligned with the user preferences, \eg, accidental clicks, or short-term intent drift. However, most existing studies utilize all available information to derive user representations, resulting in potential deviation from actual user preferences due to noise interference. Recent studies~\cite{ZhouYZW22FMLPRec, DuYZF0LS023SLIME4Rec, DuYZQZ0LS23FEARec, aaai/Shin0WP24BSARec} have attempted to address this issue by eliminating noise and emphasizing crucial information with embedding filtering techniques. Nevertheless, they are limited in considering a single sequence, focusing solely on item IDs. While DLFSRec~\cite{LiuD0P023DLFSRec} introduces a frequency-based learnable filter in the multi-sequence, it overlooks sequence-level denoising. To overcome this limitation, it is necessary to filter irrelevant signals across individual multiple sequences of item IDs and attributes.

\noindent
\textbf{(ii) Limited utilization of side-information}. 
Although intermediate fusion addresses the issue in early and late fusion strategies, it primarily focuses on utilizing item attributes to guide the importance of item IDs. Specifically, NOVA~\cite{Chang21NOVA}, DIF-SR~\cite{XieZK22DIFSR}, and ASIF~\cite{www/Wang24ASIF} exploit item attributes only for calculating attention weights. The final user representation is then obtained by aggregating the item ID vectors in the sequence. As a result, it fails to directly integrate item attributes into user representations, thereby missing strong collaborative signals across attributes.

We first employ \emph{frequency-based noise filtering} to remove noisy signals and extract salient patterns. Specifically, we transform each sequence into a frequency signal using discrete Fourier transforms. We then apply a frequency-based filtering algorithm, a common technique in digital signal processing~\cite{daglib/0078893DigitalImageProcessing, soliman1990FourierTransform, MenantNMP17DigitalSignalProcessingSurvey}. It can consider periodicity and patterns that may be difficult to discern in the time domain~\cite{ZhouYZW22FMLPRec, aaai/Shin0WP24BSARec, DuYZQZ0LS23FEARec}. Since essential information differs across item IDs and attributes, we apply frequency-based filtering to each sequence. For instance, as illustrated in Figure~\ref{fig:motivation}, the third item belongs to the ``\emph{Brand}'' of ``\emph{Apple}'', which represents a consistent pattern that should be emphasized. However, from the perspective of the category sequence, ``\emph{Earphone}'' may appear as an inconsistent pattern within the category sequence. Therefore, we employ attribute-level filtering for each sequence to identify and prioritize meaningful signals across different attributes effectively.

We then introduce \emph{dual multi-sequence fusion}, combining intermediate and early fusion. Intermediate fusion effectively aggregates \emph{ID-centric} correlation within the sequence~\cite{Chang21NOVA, XieZK22DIFSR, Lin24MSSR, www/Wang24ASIF}. As depicted in Figure~\ref{fig:motivation}, the brand and category sequences may highlight ``\emph{Apple}'' and ``\emph{Cellular phone}'', respectively. Intermediate fusion aggregates these highlighted attributes into an item ID value matrix, assigning higher attention scores to items that align with them, such as the first and last items, corresponding to ``\emph{Apple cellular phone}''. This approach ensures that the most critical attribute combinations are emphasized, allowing the model to focus on items that best represent the user's core preferences. However, it primarily captures relationships within a single attribute and may overlook broader patterns across different attributes. We thus adopt early fusion, which is more effective for identifying correlations between various attributes. For example, if a user consistently prefers the ``\emph{Apple}'' brand across different categories, early fusion can recognize this preference even when the item is not specifically highlighted in the category sequences, such as ``\emph{Apple earphone}''. Similarly, if a user prefers the ``\emph{Cellular phone}'' category regardless of brand, early fusion can effectively capture this pattern by identifying relevant items, such as ``\emph{Motorola cellular phone}'' and ``\emph{Samsung cellular phone}''. By representing items with a combination of attributes, early fusion provides a holistic view of user preferences that may not be fully captured through intermediate fusion alone. To mitigate information invasion of the na\"ive early fusion~\cite{Chang21NOVA, XieZK22DIFSR}, we also align ID and attribute representations in the same space. This allows the dual fusion approach to mitigate potential drawbacks while leveraging the strengths of both early and intermediate fusion.

To this end, we propose a novel side-information integrated sequential recommendation model, namely \textit{\textbf{D}ual Side-\textbf{I}nformation \textbf{F}iltering and \textbf{F}usion model (\textbf{DIFF})}. It consists of two key components: (i) \emph{Frequency-based Noise Filtering} and (ii) \emph{Dual Multi-sequence Fusion}. First, we remove noise and maintain only essential information based on the frequency domain. We then adjust high- and low-frequency signals for each item ID and attribute. Subsequently, filtered ID and attribute sequences are utilized in dual fusion. It consists of two distinct fusion blocks corresponding to intermediate and early fusion. As \emph{ID-centric Fusion}, intermediate fusion captures the intra-attribute correlation across items. As \emph{Attribute-enriched Fusion}, early fusion enables us to identify inter-attribute correlations across various attributes. With the proposed filtering and fusion strategy, DIFF is more robust in learning subtle and complex item relationships in multiple sequences. Experimental results show that DIFF significantly outperforms the state-of-the-art SISR models, improving performance by up to 14.1\% and 12.5\% on Recall@20 and NDCG@20 across four real-world datasets.

\begin{figure*}
\centering
\includegraphics[width=0.98\linewidth]{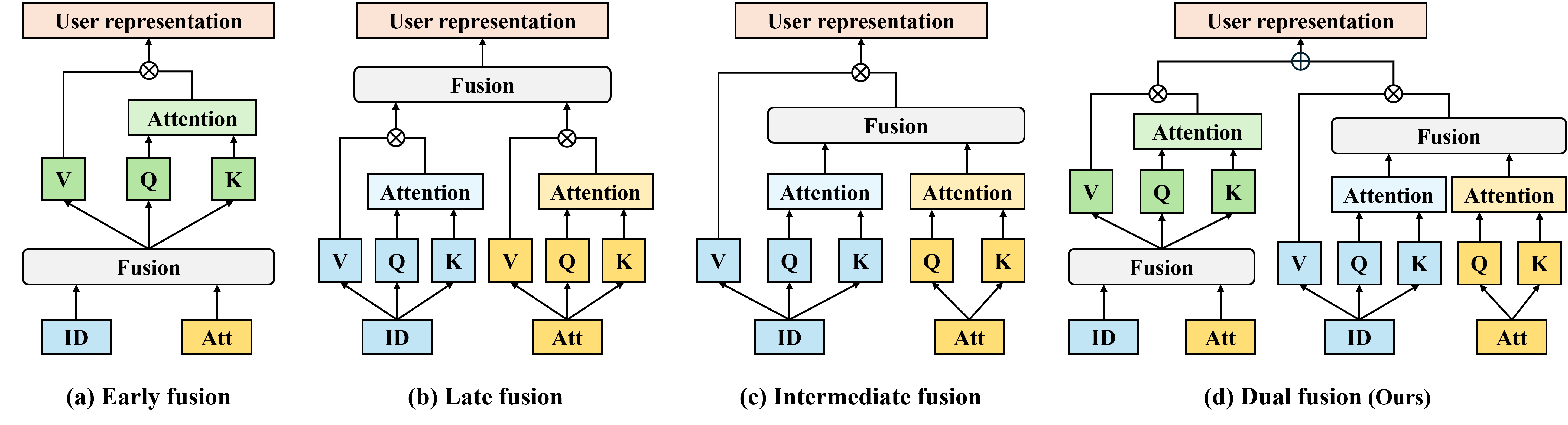} 
\vspace{-4mm}
\caption{Comparison of side information fusion methods. Existing methods are broadly categorized into (a) early, (b) late, and (c) intermediate fusion. We introduce (d) \textit{dual fusion}, which benefits from early and intermediate fusion.}\label{fig:related_work}  
\vspace{-2mm}
\end{figure*}
\section{Related Work}

\noindent
\textbf{Sequential Recommendation (SR)}. It aims to deliver the next item based on the user’s sequential interaction history. Numerous studies have employed neural architectures as encoders, \eg, Convolutional Neural Networks (CNNs)~\cite{TangW18caser}, Recurrent Neural Networks (RNNs)~\cite{LiRCRLM17NARM, HidasiKBT15GRU4Rec}, Graph Neural Networks (GNNs)~\cite{WuT0WXT19SRGNN, GuptaGMVS19NISER}, and transformers~\cite{KangM18SASRec, SunLWPLOJ19BERT4Rec, kdd/MaKL19HGN}. Recently, some studies~\cite{ZhouYZW22FMLPRec, DuYZF0LS023SLIME4Rec, DuYZQZ0LS23FEARec, aaai/Shin0WP24BSARec} have shifted from the time domain to the frequency domain, identifying salient patterns in user behavior. However, they primarily focus on learning item correlations only with item ID sequences, neglecting side-information that provides a rich context for user behavior.

\vspace{2mm}
\noindent
\textbf{Side-Information Integrated SR (SISR)}. It utilizes both item IDs and attributes in the user’s sequential history. S$^{3}$-Rec~\cite{ZhouWZZWZWW20S3Rec} adopts self-supervised auxiliary tasks to learn the relationship between item IDs and attributes. DLFSRec~\cite{LiuD0P023DLFSRec} utilizes a distribution-based learnable filter, representing ID and attributes by Gaussian distribution to capture their uncertainty. Then, various fusion methods~\cite{ZhangZLSXWLZ19FDSA, Chang21NOVA, XieZK22DIFSR, Lin24MSSR, www/Wang24ASIF} have been proposed to combine item IDs and attributes in the self-attention mechanism. As depicted in Figure~\ref{fig:related_work}, they can be categorized into three pillars as follows~\cite{mms/AtreyHEK10MMsurvey, pami/BaltrusaitisAM19MMsurvey}. 

\begin{itemize}[leftmargin=*,topsep=0pt,itemsep=-1ex,partopsep=1ex,parsep=1ex]
\item \textbf{Early fusion}: It incorporates item ID and attributes at the input level as illustrated in Figure~\ref{fig:related_work}(a). GRU4Rec$_{\text{F}}$ and SASRec$_{\text{F}}$~\cite{ZhouWZZWZWW20S3Rec} create a unified representation by combining sequences of IDs and attributes as input before feeding it into the model by concatenation, summation, or gating. Although they combine the ID-attribute interactions via fused embeddings, it is challenging to learn the entangled embedding space of IDs and attributes as pointed out in the previous work, \ie, information invasion~\cite{Chang21NOVA}.

\item \textbf{Late fusion}: It delays the integration of item IDs and attributes until the model’s final layer as in Figure~\ref{fig:related_work}(b). FDSA~\cite{ZhangZLSXWLZ19FDSA} adopts separated self-attention blocks to encode item IDs and attributes independently. While it can capture the different contexts of individual item sequences, it risks missing out on interactions between IDs and attributes. 

\item \textbf{Intermediate fusion}: It considers the interaction between item IDs and attributes in the intermediate layer as shown in Figure~\ref{fig:related_work}(c). It first extracts meaningful patterns from each sequence before combining them. Concretely, NOVA~\cite{Chang21NOVA} DIF-SR~\cite{XieZK22DIFSR}, ASIF~\cite{www/Wang24ASIF} integrate ID and attributes at an intermediate layer to calculate the query and key matrices in the self-attention block. However, they utilize attributes to obtain attention scores, overlooking direct correlations across attributes.
\end{itemize}

Under the categorization above, some methods adopt the combination of intermediate and late fusion. ESIF~\cite{ZhengS24ESIF} aggregates intermediate fused attention utilizing each attribute value matrix, MSSR~\cite{Lin24MSSR} introduces both intra-sequence and inter-sequence attention to consider the correlation between item ID and attribute sequences. However, existing studies do not explicitly combine the advantages of early and intermediate fusion.

\section{Preliminaries}\label{sec:preliminary}

\noindent
\textbf{Problem Formulation}. Let $\mathcal{I} = \{i_1, \dots, i_n\}$ represent a set of $n$ items. The user's item sequence is denoted as $s = [i_{1}, \dots, i_{|s|}]$, where $i_j$ is the $j$-th item in the sequential order, and $|s|$ is the total number of items interacted with by the user. Following the previous studies~\cite{ZhangZLSXWLZ19FDSA, Lin24MSSR, www/Wang24ASIF}, we mainly consider item-related side-information, \eg, brand and category. For side-information integrated sequential recommendation, each item $i \in \mathcal{I}$ is described by its unique item ID and multiple attributes. Specifically, it is represented as $i_j = \{v_j, a_{1,j}, \dots, a_{m,j}\}$, where $v_j$ is its item ID, $a_{k,j}$ is the $k$-th attribute type, and $m$ is the total number of attributes. Our goal is to predict the next item the user is most likely to prefer, expressed as $\text{argmax}_{j \in \mathcal{I}} P(i_{|s|+1} = j \ | \ s)$.

\begin{figure*}
\centering
\includegraphics[width=0.99\linewidth]{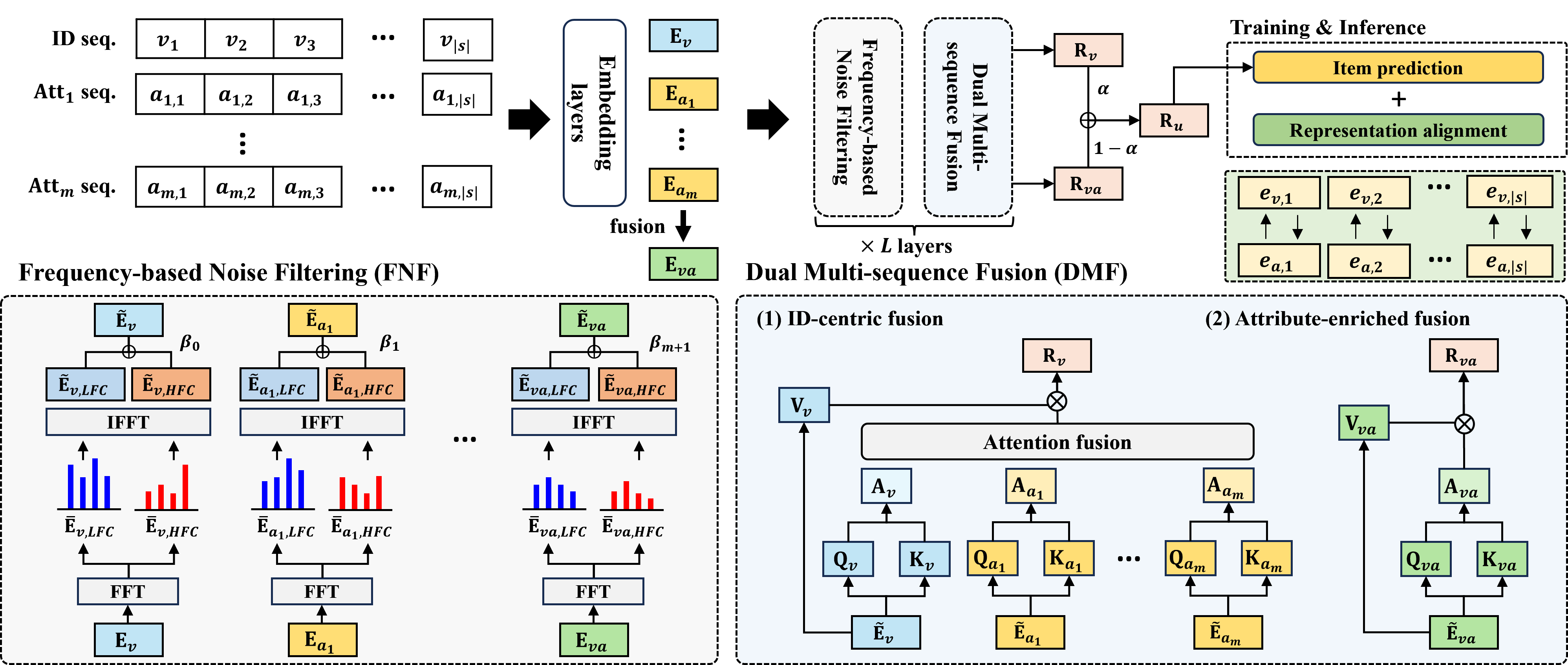}
\vspace{-1mm}
\caption{An overview of DIFF. DIFF processes both independent sequences and early fused sequences via $L$ layers of two components: (i) Frequency-based Noise Filtering and (ii) Dual Multi-sequence Fusion. DIFF yields filtered user representations that fully integrates item attributes. Multi-task learning with representation alignment ensures smooth ID-attribute fusion.}\label{fig:model}
\vspace{-2mm}
\end{figure*}

\vspace{1mm}
\noindent
\textbf{Discrete Fourier Transform (DFT)}. The DFT is a fundamental component of digital signal processing, converting a sequence in the time domain into the frequency domain. Given a sequence with length $N$, the DFT is represented as $\mathcal{F}: \mathbb{R}^{N} \to \mathbb{C}^{N}$, and its inverse, \ie, the inverse discrete Fourier transform (IDFT), is denoted as $\mathcal{F}^{-1}: \mathbb{C}^{N} \to \mathbb{R}^{N}$. The DFT can be performed by multiplying a sequence matrix $\mathbf{X} \in  \mathbb{R}^{N \times d}$ by the matrix $\mathbf{F} \in \mathbb{C}^{N \times N}$.
\begin{equation} \label{eq:dft}
    \bar{\mathbf{X}} = \mathcal{F}(\mathbf{X}) = \mathbf{F} \mathbf{X} = \frac{1}{\sqrt{N}}
    \begin{bmatrix}
    1 & 1 & \cdots & 1 \\
    1 & e^{\frac{-2 \pi i}{N}} & \cdots & e^{\frac{-2 \pi i(N-1)}{N}} \\
    \vdots & \vdots & \vdots & \vdots \\
    1 & e^{\frac{-2 \pi i(N-1)}{N}} & \cdots & e^{\frac{-2 \pi i(N-1)^{2}}{N}} \\
    \end{bmatrix} \mathbf{X},
\end{equation}
% 원본
where $i$ is the imaginary unit, and $\bar{\mathbf{X}} \in \mathbb{C}^{N \times d}$ is the frequency component of sequence $\mathbf{X}$.
Interestingly, $\bar{\mathbf{X}}$ can be separated into two parts: \emph{low-frequency} and \emph{high-frequency} components. We define the first $c$ rows as a low-frequency component $\bar{\textbf{X}}_{LFC} \in \mathbb{C}^{c \times d}$ and the remaining rows as a high-frequency component $\bar{\textbf{X}}_{HFC} \in \mathbb{C}^{(N-c) \times d}$. IDFT is then applied to convert each component into a different signal type.
\begin{equation} \label{eq:fc}
\begin{gathered}
    \tilde{\mathbf{X}}_{LFC} = \mathcal{F}^{-1}(\bar{\mathbf{X}}_{LFC}) = [\mathbf{f}_{1}^{*\top}, \dots ,\mathbf{f}_{c}^{*\top}]\bar{\mathbf{X}}_{LFC}, \\
\tilde{\mathbf{X}}_{HFC} = \mathcal{F}^{-1}(\bar{\mathbf{X}}_{HFC}) = [\mathbf{f}_{c+1}^{*\top},\dots ,\mathbf{f}_{N}^{*\top}]\bar{\mathbf{X}}_{HFC},
\end{gathered}
\end{equation}

\noindent
where $\mathbf{f}_i$ represents the $i$-th row vector in the matrix $\mathbf{F}$, and `$*$' denotes the conjugate operation. The low-frequency component $\tilde{\mathbf{X}}_{LFC} \in \mathbb{R}^{N \times d}$ captures the overall trend of the sequence, representing the signal that does not change frequently. In contrast, the high-frequency component $\tilde{\mathbf{X}}_{HFC} \in \mathbb{R}^{N \times d}$ represents the signal with abrupt variations. Note that we utilize Fast Fourier Transform (FFT)~\cite{CooleyJ65Fourier,FrigoJ05FFT}, an efficient algorithm computing the DFT and IDFT.

\section{Proposed Model: DIFF}\label{sec:model}

In this section, we present the \emph{\textbf{D}ual Side-\textbf{I}nformation \textbf{F}iltering and \textbf{F}usion (\textbf{DIFF})} model, which effectively removes noisy signals and fully leverages the correlation across item IDs and attributes. Figure~\ref{fig:model} depicts the overall architecture of DIFF, which consists of two main components: (i) \emph{Frequency-based Noise Filtering} and (ii) \emph{Dual Multi-sequence Fusion}. Specifically, frequency-based noise filtering is used to eliminate noise and extract essential signals (Section~\ref{sec:fnf_layer}). Subsequently, dual multi-sequence fusion is employed to learn complex interactions across filtered item ID and attribute sequences (Section~\ref{sec:hmf_layer}). We also adopt an alignment loss to prevent information invasion between item IDs and attributes (Section~\ref{sec:alignment}). Lastly, we explain the training and inference of DIFF (Section~\ref{sec:train_test}).

\subsection{Frequency-based Noise Filtering}\label{sec:fnf_layer}

We employ \emph{frequency-based noise filtering} to reduce irrelevant variations and distinguish essential patterns associated with consistent user preferences. The item sequence is converted to a frequency signal using the Fourier transform. Since item IDs and attributes exhibit different patterns, frequency-based filtering is applied independently to the item ID and attribute sequences.

\vspace{1mm}
\noindent
\textbf{Embedding Layer}. Given a user sequence $s =[i_{1}, i_{2}, \dots, i_{|s|}]$, we first obtain the embedding matrices for item ID sequence and attribute sequences. 
\begin{equation}
\begin{gathered}
    \mathbf{E}_{v} = \mathcal{E}_{v}(v_{1},v_{2},\dots,v_{|s|}), \\
    \mathbf{E}_{a_{k}} = \mathcal{E}_{a_{k}}(a_{k,1},a_{k,2},\dots,a_{k,|s|}) \ \forall k \in [1, m],
\end{gathered}
\end{equation}
where $\mathbf{E}_{v}$ and $ \mathbf{E}_{a_{k}} \in \mathbb{R}^{|s| \times d}$ are the resulting embedding matrices for the item ID sequence and the $k$-th attribute sequence, respectively. Also, $\mathcal{E}_{v}$ and $\mathcal{E}_{a_{k}}$ are embedding layers for the item ID and $k$-th item attribute, respectively. 

While existing studies~\cite{XieZK22DIFSR, Lin24MSSR} have primarily focused on optimizing the intermediate fusion, our approach considers both early and intermediate fusion to capture essential patterns through integrated embeddings across item IDs and attributes. To achieve this, we obtain a fused embedding $\mathbf{E}_{va}$ for early fusion that combines the item ID and all attributes:
\begin{equation}
    \mathbf{E}_{va} = \text{Fusion}\left (\mathbf{E}_{v},\mathbf{E}_{a_{1}},\dots,\mathbf{E}_{a_{m}}\right),
\end{equation}
where $\mathbf{E}_{va} \in \mathbb{R}^{|s| \times d}$ and $\text{Fusion}(\cdot)$ denotes the fusion function for item ID and attribute embeddings. Following the prior studies~\cite{Chang21NOVA, XieZK22DIFSR, Lin24MSSR, www/Wang24ASIF}, various fusion functions can be used, \ie, summation, concatenation, or gating.  

\vspace{1mm}
\noindent
\textbf{Frequency-based Filtering}. We employ the filtering method to remove noise and spurious signals for each sequence. As pointed out in previous studies~\cite{www/Wang24ASIF, LiuD0P023DLFSRec}, it is crucial to enhance the utilization of side-information by alleviating noisy interference. To achieve this, we utilize the discrete Fourier transform to project a sequence into the frequency domain. Specifically, we define the low- and high-frequency components of item ID embeddings as $\bar{\mathbf{E}}_{v,LFC} \in \mathbb{C}^{c \times d}$ and $\bar{\mathbf{E}}_{v,HFC} \in \mathbb{C}^{(|s|-c) \times d}$, respectively. 
\begin{equation}
\begin{gathered}
    \tilde{\mathbf{E}}_{v, LFC} = \mathcal{F}^{-1}(\bar{\mathbf{E}}_{v,LFC}), \\
    \tilde{\mathbf{E}}_{v, HFC} = \mathcal{F}^{-1}(\bar{\mathbf{E}}_{v,HFC}).
\end{gathered}
\end{equation}
$\tilde{\mathbf{E}}_{v, LFC}, \tilde{\mathbf{E}}_{v, HFC} \in \mathbb{R}^{|s| \times d}$ represent the low- and high-frequency components of the item ID embedding, respectively. Similarly, we obtain the low- and high-frequency components for each attribute embedding and fused embedding through frequency-based filtering.
\begin{equation}
\begin{gathered}
\tilde{\mathbf{E}}_{a_{k}, LFC} = \mathcal{F}^{-1}(\bar{\mathbf{E}}_{a_{k}, LFC}), \ \ \forall k \in [1, m], \\
    \tilde{\mathbf{E}}_{a_{k}, HFC} = \mathcal{F}^{-1}(\bar{\mathbf{E}}_{a_{k}, HFC}), \ \ \forall k \in [1, m],
\end{gathered}
\end{equation}

\begin{equation}
\begin{gathered}
\tilde{\mathbf{E}}_{va, LFC} = \mathcal{F}^{-1}(\bar{\mathbf{E}}_{va, LFC}), \\
    \tilde{\mathbf{E}}_{va, HFC} = \mathcal{F}^{-1}(\bar{\mathbf{E}}_{va, HFC}),
\end{gathered}
\end{equation}

\noindent
where $\tilde{\mathbf{E}}_{a_{k}, LFC}, \tilde{\mathbf{E}}_{a_{k}, HFC}, \tilde{\mathbf{E}}_{va, LFC}, \tilde{\mathbf{E}}_{va, HFC} \in \mathbb{R}^{|s| \times d}$.

From a frequency perspective, low-frequency signals represent stable patterns that change minimally over a sequence, while high-frequency signals exhibit rapid fluctuations. In the context of item sequences, the low-frequency component can be interpreted as representing long-term and consistent user interests. In contrast, the high-frequency component reflects short-term and volatile interests. While user's long-term consistent interests are crucial for making accurate recommendations, short-term interests that emerge suddenly are often less significant and may serve as noisy information. 

To prioritize long-term stable user interests, we derive the filtered embeddings $\tilde{\mathbf{E}}_{v}$, $\tilde{\mathbf{E}}_{a_{k}}$, and $\tilde{\mathbf{E}}_{va}$ for each sequence by adjusting the impact of the high-frequency component.
\begin{equation}
\begin{gathered}
    \tilde{\mathbf{E}}_{v} = \tilde{\mathbf{E}}_{v,LFC} + \beta_{0} \tilde{\mathbf{E}}_{v,HFC}, \\
    \tilde{\mathbf{E}}_{a_{k}} = \tilde{\mathbf{E}}_{a_{k},LFC} + \beta_{k} \tilde{\mathbf{E}}_{a_{k},HFC}, \ \forall k \in [1, m], \\
    %    \vdots \\
    %\tilde{\mathbf{E}}_{a_{m}} = \mathbf{E}_{a_{m},LFC} + \beta_{m} \mathbf{E}_{a_{m},HFC}, \\
    \tilde{\mathbf{E}}_{va} = \tilde{\mathbf{E}}_{va,LFC} + \beta_{m+1} \tilde{\mathbf{E}}_{va,HFC},
\end{gathered}
\end{equation}
where $\beta_{0},\beta_{1},...,\beta_{m+1}$ are trainable scalar parameters used to adjust the high-frequency components of each input embedding. Empirically, we observe that $\beta$ is trained to a very small value, \ie, the impact of short-term fluctuating interests is reduced.

\subsection{Dual Multi-sequence Fusion}\label{sec:hmf_layer}

We leverage both early and intermediate fusion to fully exploit the potential of side-information. Since two fusion strategies can capture different correlations across items and attributes, our dual fusion can be more effective than solely relying on intermediate fusion~\cite{Chang21NOVA,XieZK22DIFSR,www/Wang24ASIF}. 
Specifically, early fusion effectively captures inter-attribute correlations, while intermediate fusion focuses on intra-attribute correlations within individual attributes.

\vspace{1mm}
\noindent
\textbf{ID-centric Fusion}. 
We employ ID-centric fusion to better capture the correlation between item IDs. This approach, a form of intermediate fusion, is also utilized in existing studies~\cite{XieZK22DIFSR}. We project ID and attribute embedding sequences onto different query and key matrices. The query and key matrices for the $h$-th attention head are as follows:
\begin{equation}
\begin{gathered}
    \mathbf{Q}_{v}^{h}=\tilde{\mathbf{E}}_{v} \mathbf{W}_{Q,v}^{h} , \mathbf{K}_{v}^{h}=\tilde{\mathbf{E}}_{v} \mathbf{W}_{K,v}^{h} , \\    %\mathbf{Q}_{a_{1}}^{h}=\tilde{\mathbf{E}}_{a_{1}}\mathbf{W}_{Q,a_{1}}^{h}, \\
    %\vdots \\
    \mathbf{Q}_{a_{k}}^{h}=\tilde{\mathbf{E}}_{a_{k}}\mathbf{W}_{Q,a_{k}}^{h}, \mathbf{K}_{a_{k}}^{h}=\tilde{\mathbf{E}}_{a_{k}}\mathbf{W}_{K,a_{k}}^{h}, \ \forall k \in [1, m],
\end{gathered}     
\end{equation}
where $\mathbf{W}_{Q,v}^{h}, \mathbf{W}_{K,v}^{h} \in \mathbb{R}^{d \times d_{h}}$ are query and key projection matrices for item IDs, and $\mathbf{W}_{Q,a_{k}}^{h}, \mathbf{W}_{K,a_{k}}^{h} \in \mathbb{R}^{d \times d_{h}}$ are query and key projection matrices for the $k$-th item attribute. We then compute the attention score for each sequence via the dot-product of the query-key pairs.
\begin{equation}
\begin{gathered}
    \mathbf{A}_{v}^{h}=\mathbf{Q}_{v}^{h}\left(\mathbf{K}_{v}^{h} \right)^{\top}, \\
    %\vdots \\
    \mathbf{A}_{a_{k}}^{h}=\mathbf{Q}_{a_{k}}^{h}\left(\mathbf{K}_{a_{k}}^{h} \right)^{\top}, \ \forall k \in [1, m],
\end{gathered}  
\end{equation}
where $\mathbf{A}_{v}^{h} \in \mathbb{R}^{|s| \times |s|}$ and $\mathbf{A}_{a_{k}}^{h} \in \mathbb{R}^{|s| \times |s|}$ denote attention score matrices of ID sequence and the $k$-th attribute sequence obtained from $h$-th attention head. Finally, we fuse the item correlations from item IDs and attributes, \ie, attention score matrices, and aggregate them into an item ID value matrix.
\begin{equation}
    \begin{gathered}  
    \mathbf{R}_{v} = \text{FFN}(\text{concat}(\mathbf{R}_{v}^{1},\dots,\mathbf{R}_{v}^{H})\mathbf{W}_{v}), \\
    \text{where} \ \mathbf{R}_{v}^{h} = \text{softmax}\left(\frac{\text{Fusion}\left(\mathbf{A}_{v}^{h},\cdots,\mathbf{A}_{a_{m}}^{h} \right)}{\sqrt{d_{h}}}\right)\mathbf{V}_{v}^{h}.
    \end{gathered}  
\end{equation}
Here, $H$ is the number of attention heads and $\mathbf{V}_{v}^{h} = \tilde{\mathbf{E}}_{v}\mathbf{W}_{V,v}^{h}$ denotes the value matrix of item IDs at the $h$-th attention head. $\mathbf{W}_{v} \in \mathbb{R}^{d \times d}$ is a weight parameter matrix, $\text{concat}(\cdot)$ and $\text{FFN}(\cdot)$ indicate concatenation and feed-forward network, respectively.

\vspace{1mm}
\noindent
\textbf{Attribute-enriched Fusion}. We adopt attribute-enriched fusion to reflect the inter-attribute correlations across various attributes, \ie, early fusion. Specifically, we apply self-attention to the fused embeddings as follows.
\begin{equation}    
    \mathbf{Q}_{va}^{h}=\tilde{\mathbf{E}}_{va}\mathbf{W}_{Q,va}^{h} \ ,\ \mathbf{K}_{va}^{h}=\tilde{\mathbf{E}}_{va}\mathbf{W}_{K,va}^{h} \ , 
\end{equation}
where $\mathbf{W}_{Q,va}^{h}, \mathbf{W}_{K,va}^{h} \in \mathbb{R}^{d \times d_{h}}$ are query, key projection matrices for the fused sequence, respectively. The attention score is computed using the query and key matrices of the fused sequence, thereby explicitly modeling strong correlations across item IDs and attributes. 
\begin{equation}
    \mathbf{A}_{va}^{h}=\mathbf{Q}_{va}^{h}\left (\mathbf{K}_{va}^{h}\right)^{\top},
\end{equation}
where $\mathbf{A}_{va}^{h} \in \mathbb{R}^{|s| \times |s|}$ is an attention score matrix of the fused embedding sequence for $h$-th attention head. We then derive the user representation from these fused embeddings.
\begin{equation}
    \begin{gathered} 
    \mathbf{R}_{va} = \text{FFN}(\text{concat}(\mathbf{R}_{va}^{1}, \dots, \mathbf{R}_{va}^{H})\mathbf{W}_{va}), \\
    \text{where} \ \mathbf{R}_{va}^{h} = \text{softmax}\left(\frac{\mathbf{A}_{va}^{h}}{\sqrt{d_{h}}}\right)\mathbf{V}_{va}^{h},
    \end{gathered}
\end{equation}
where $\mathbf{V}_{va}^{h} \in \mathbb{R}^{d \times d_{h}}$ denotes a projected value matrix for the fused embeddings. $\mathbf{W}_{va} \in \mathbb{R}^{d \times d}$ is a weight parameter matrix.
% Note that the self-attention is performed upon noise-filtered embedding of IDs and attributes.

\noindent
\textbf{User Representation}. The final user representation is obtained by aggregating early and intermediate fusion results. While the ID-centric representation via intermediate fusion emphasizes fine-grained interactions at the individual item level, the attribute-enriched representation via early fusion explicitly captures strong attribute correlations. The final representation is computed as:
\begin{equation}
    \mathbf{R}_{u}= \alpha \mathbf{R}_{v} + (1-\alpha) \mathbf{R}_{va},
\end{equation}
where $\mathbf{R}_{u} \in \mathbb{R}^{|s| \times d}$ and $\alpha$ denotes the representation aggregating hyperparameter. The last element in $\mathbf{R}_{u}$, \ie, $\mathbf{r}_{u, |s|} \in \mathbb{R}^{d}$, is used as the user representation vector for prediction.

\subsection{Representation Alignment}\label{sec:alignment}
Item IDs and attributes are initially embedded in separate spaces. However, they need to be semantically consistent since both ID-centric and attribute-enriched representations are used for the final user representation. For that, we leverage a contrastive loss to align the embedding spaces of item IDs and fused attributes. Inspired by previous work~\cite{www/Wang24ASIF}, we align the similarity between the item ID and the fused attribute embedding vectors.
\begin{equation}
    \begin{gathered} 
    \mathbf{\hat{Y}}_{v,a}=\text{softmax}\left(\frac{\mathbf{E}_{v} \mathbf{E}_{a}^{\top}}{\tau}\right), \mathbf{\hat{Y}}_{a,v}=\text{softmax}\left(\frac{\mathbf{E}_{a} \mathbf{E}_{v}^{\top}}{\tau}\right), \\
    \text{where} \ \mathbf{E}_{a} = \text{Fusion}(\mathbf{E}_{a_{1}},\dots,\mathbf{E}_{a_{m}}).
    \end{gathered}
\end{equation}
Here, $\mathbf{E}_{a} \in \mathbb{R}^{|s| \times d}$ is a fused embedding matrix of item attributes with the fusion function. For that, the summation function is used. In this process, we use normalized item ID and fused attribute embeddings for stable training. The learnable temperature $\tau$ is used as a scaling factor. The final alignment loss is defined as follows:
\begin{equation}
    \mathcal{L}_{align}=-\frac{1}{2b}\sum_{i=1}^{b}\sum\left(\mathbf{Y}^{i} \odot \log\hat{\mathbf{Y}}^{i}_{v,a} + \mathbf{Y}^{i} \odot \log\hat{\mathbf{Y}}^{i}_{a,v}\right),
\end{equation}
where $\odot$ denotes element-wise product, $\mathbf{Y}^{i} \in \{0,1\}^{|s| \times |s|}$ is the ground truth of the $i$-th sequence, and $b$ is the number of sequences in the mini-batch. Each element of $\mathbf{Y}^{i}$ is defined as follows.
\begin{equation}
\mathbf{Y}^{i}_{j,k} = \begin{cases}
                        1 & \text{if } \mathbf{E}_{a}^{j} = \mathbf{E}_{a}^{k} \\
                        0 & \text{otherwise} \\
                        \end{cases}
                                    \text{ for } j, k \in \{1, \dots, |s|\},
\end{equation}
where $\mathbf{E}_{a}^{j}$ and $\mathbf{E}_{a}^{k}$ are the fused attribute embedding vectors obtained from the $j$-th and $k$-th items in $i$-th sequence, respectively.

\subsection{Training and Inference}\label{sec:train_test}

For inference, we make predictions using the final user representation vector $\mathbf{r}_{u, |s|}$ and the item ID embedding matrix $\mathbf{E}$.
\begin{equation} \label{eq:item_pred}
    \hat{\mathbf{y}} = \text{softmax}(\mathbf{r}_{u, |s|} \mathbf{E}^{\top}),
\end{equation}

\noindent
where $\hat{\mathbf{y}} \in \mathbb{R}^{n}$. To calculate the recommendation loss, we employ the cross-entropy loss function.
\begin{equation} \label{eq:rec_loss}
    \mathcal{L}_{rec} = - \frac{1}{b}\sum_{i=1}^{b}\mathbf{y}^{(i)}\log\hat{\mathbf{y}}^{(i)},
\end{equation}
where $\mathbf{y}^{(i)} \in \{0, 1\}^{n}$ is the one-hot encoded ground truth vector of the $i$-th sequence in the mini-batch, with the element corresponding to the target item set to 1 and all others to 0. 

Finally, we train our model by combining the recommendation loss and representation alignment loss.
\begin{equation}
    \mathcal{L} = \mathcal{L}_{rec} + \lambda \mathcal{L}_{align},
\end{equation}

\noindent
where $\lambda$ is the hyperparameter to control the loss $\mathcal{L}_{align}$.

\section{Experimental Setup}\label{sec:setup}

\begin{table}[]\small
\caption{Data statistics after preprocessing. Avg. Length indicates the average number of interactions per user.}  \label{tab:dataset}
\vspace{-2mm}
\begin{tabular}{c|cccc}
\toprule
Dataset         & Yelp    & Beauty  & Sports  & Toys    \\ \midrule
\# Users        & 30,449  & 22,363  & 35,598  & 19,412  \\
\# Items        & 20,068  & 12,101  & 18,357  & 11,924  \\
\# Interactions & 317,182 & 198,502 & 296,337 & 167,597 \\
Avg. Length     & 10.4    & 8.9     & 8.3     & 8.6     \\ 
Sparsity        & 99.95\% & 99.93\% & 99.95\% & 99.93\% \\
\bottomrule
\end{tabular}
\end{table}

\noindent
\textbf{Datasets}. We conduct extensive experiments on four real-world datasets following \cite{XieZK22DIFSR, Lin24MSSR}, \ie, Yelp~\footnote{\url{https://www.yelp.com/dataset}} and Amazon review dataset\\~\cite{sigir/McAuleyTSH15Amazonreview}~\footnote{\url{https://jmcauley.ucsd.edu/data/amazon/}}. \textbf{Yelp} is a well-known business recommendation dataset. The attributes of categories, cities, and positions are utilized as side-information. We select three widely used subcategories that are constructed from the Amazon review datasets: \textbf{Beauty}, \textbf{Sports}, and \textbf{Toys}. They consist of item metadata and reviews collected from 1996 to 2014, and we utilize the categories, brands, and positions as side-information. As in the previous works~\cite{XieZK22DIFSR, Lin24MSSR}, we use the 5-core setting, which removes users and items that occur less than five times. The detailed statistics for the pre-processed datasets are shown in Table~\ref{tab:dataset}.
%%%%%%%%%%%%%%%% @10, 20 version (No @5)
\begin{table*}[] 
\small
\centering
\setlength{\tabcolsep}{2.2pt}
\caption{Overall performance comparison on four datasets. * denotes that DIFF shows statistically significant improvement $(p < 0.05)$ over the best competitive model. The best results are marked in bold, and the second best results are \ul{underlined}.} \label{tab:overall_performance}
\vspace{-3mm}
% \begin{tabular}{c|c|ccccc|cccccccc|c}
\begin{tabular}{>{\centering\arraybackslash}p{0.9cm}|>{\centering\arraybackslash}p{0.8cm}|>{\centering\arraybackslash}p{0.8cm} >{\centering\arraybackslash}p{0.8cm} >{\centering\arraybackslash}p{1.05cm} >{\centering\arraybackslash}p{0.9cm}|>{\centering\arraybackslash}p{1.1cm}>{\centering\arraybackslash}p{1.1cm} >{\centering\arraybackslash}p{0.8cm} >{\centering\arraybackslash}p{0.9cm} >{\centering\arraybackslash}p{0.8cm} >{\centering\arraybackslash}p{0.8cm} >{\centering\arraybackslash}p{0.9cm} >{\centering\arraybackslash}p{0.8cm} >{\centering\arraybackslash}p{0.8cm} |>{\centering\arraybackslash}p{1cm}|>{\centering\arraybackslash}p{0.8cm}}

\toprule
\multirow{2}{*}{Dataset} & \multirow{2}{*}{Metric} & \multicolumn{4}{c|}{SR baselines} & \multicolumn{9}{c|}{SISR baselines} & \multirow{2}{*}{} &  \\ 
 &  & SASRec & DuoRec & FMLPRec & BSARec & GRU4Rec$_{\text{F}}$ & SASRec$_{\text{F}}$ & S$^3$-Rec & DLFSRec & FDSA & NOVA & DIF-SR & MSSR & ASIF & DIFF & Gain \\ \midrule
\multirow{4}{*}{Yelp} 
 & R@10 & 0.0607 & 0.0631 & 0.0711 & 0.0701 & 0.0414 & 0.0435 & 0.0598 & 0.0551 & 0.0537 & 0.0614 & 0.0686 & 0.0712 & \ul{0.0724} & \textbf{0.0815*} & 12.5\% \\
 & R@20 & 0.0875 & 0.0909 & 0.1029 & 0.1023 & 0.0679 & 0.0706 & 0.0869 & 0.0857 & 0.0856 & 0.0886 & 0.0998 & 0.1040 & \ul{0.1052} & \textbf{0.1200*} & 14.1\% \\
 & N@10 & 0.0383 & 0.0385 & 0.0424 & 0.0423 & 0.0213 & 0.0225 & 0.0377 & 0.0312 & 0.0284 & 0.0384 & 0.0415 & 0.0425 & \ul{0.0427} & \textbf{0.0470*} & 10.2\% \\
 & N@20 & 0.0451 & 0.0455 & 0.0506 & 0.0503 & 0.0280 & 0.0293 & 0.0445 & 0.0388 & 0.0364 & 0.0452 & 0.0493 & 0.0507 & \ul{0.0510} & \textbf{0.0567*} & 11.1\% \\ \midrule
\multirow{4}{*}{Beauty} 
 & R@10 & 0.0842 & 0.0865 & 0.0855 & 0.0871 & 0.0682 & 0.0804 & 0.0839 & 0.0774 & 0.0811 & 0.0817 & 0.0891 & 0.0883 & \ul{0.0920} & \textbf{0.0935*} & 1.6\% \\
 & R@20 & 0.1191 & 0.1225 & 0.1239 & 0.1260 & 0.0991 & 0.1123 & 0.1186 & 0.1217 & 0.1152 & 0.1169 & 0.1281 & 0.1256 & \ul{0.1322} & \textbf{0.1347*} & 1.9\% \\
 & N@10 & 0.0424 & 0.0448 & 0.0426 & 0.0437 & 0.0380 & \ul{0.0468} & 0.0420 & 0.0337 & 0.0461 & 0.0415 & 0.0444 & 0.0454 & 0.0463 & \textbf{0.0526*} & 12.5\% \\
 & N@20 & 0.0511 & 0.0538 & 0.0522 & 0.0535 & 0.0458 & 0.0549 & 0.0508 & 0.0448 & 0.0547 & 0.0504 & 0.0542 & 0.0548 & \ul{0.0564} & \textbf{0.0632*} & 12.0\% \\ \midrule
\multirow{4}{*}{Sports}
 & R@10 & 0.0487 & 0.0489 & 0.0495 & 0.0506 & 0.0410 & 0.0443 & 0.0465 & 0.0402 & 0.0498 & 0.0473 & 0.0534 & 0.0549 & \ul{0.0568} & \textbf{0.0574} & 1.1\% \\
 & R@20 & 0.0709 & 0.0723 & 0.0743 & 0.0741 & 0.0625 & 0.0648 & 0.0677 & 0.0656 & 0.0723 & 0.0690 & 0.0784 & 0.0809 & \ul{0.0827} & \textbf{0.0853*} & 3.2\% \\
 & N@10 & 0.0231 & 0.0246 & 0.0232 & 0.0239 & 0.0218 & 0.0251 & 0.0226 & 0.0183 & \ul{0.0282} & 0.0229 & 0.0251 & 0.0261 & 0.0268 & \textbf{0.0310*} & 10.1\% \\
 & N@20 & 0.0287 & 0.0305 & 0.0295 & 0.0298 & 0.0272 & 0.0302 & 0.0279 & 0.0246 & \ul{0.0339} & 0.0283 & 0.0314 & 0.0326 & 0.0333 & \textbf{0.0381*} & 12.5\% \\ \midrule
\multirow{4}{*}{Toys} 
 & R@10 & 0.0889 & 0.0939 & 0.0923 & 0.0928 & 0.0643 & 0.0789 & 0.0913 & 0.0820 & 0.0884 & 0.0930 & 0.1011 & \ul{0.1020} & 0.1007 & \textbf{0.1023} & 0.3\% \\
 & R@20 & 0.1225 & 0.1287 & 0.1302 & 0.1293 & 0.0950 & 0.1112 & 0.1238 & 0.1260 & 0.1221 & 0.1253 & 0.1379 & \ul{0.1405} & 0.1393 & \textbf{0.1425*} & 1.3\% \\
 & N@10 & 0.0436 & 0.0481 & 0.0446 & 0.0460 & 0.0350 & 0.0456 & 0.0449 & 0.0364 & 0.0506 & 0.0458 & 0.0504 & \ul{0.0510} & 0.0496 & \textbf{0.0553*} & 8.6\% \\
 & N@20 & 0.0521 & 0.0569 & 0.0541 & 0.0552 & 0.0427 & 0.0537 & 0.0531 & 0.0475 & 0.0591 & 0.0539 & 0.0597 & \ul{0.0607} & 0.0593 & \textbf{0.0656*} & 8.1\% \\ \bottomrule

\end{tabular}
\end{table*}

\vspace{1mm}
\noindent
\textbf{Evaluation Protocols and Metrics}. Following~\cite{XieZK22DIFSR, Lin24MSSR}, we adopt the \textit{leave-one-out} strategy to split train, validation, and test sets. For each user sequence, we use the last item for testing, the second last item for validation, and the rest items for training. All models are evaluated in a \textit{full ranking} scenario on all items rather than sampled items following~\cite{XieZK22DIFSR, Lin24MSSR}. We opt not to penalize repeated items unlike~\cite{LiuD0P023DLFSRec,aaai/Shin0WP24BSARec} that have previously appeared within the user history to maintain consistency across diverse datasets. Applying such penalties can negatively impact models on datasets like Yelp, where repeated interactions are common, leading to biased performance estimation. For evaluation metrics, we employ top-\textit{k} Recall (R@\textit{k}) and top-\textit{k} Normalized Discounted Cumulative Gain (N@\textit{k}) with $k=\{10, 20\}$.

\vspace{1mm}
\noindent
\textbf{Baselines}. We thoroughly compare our proposed method with two categories: sequential recommendation (SR) and side-information integrated sequential recommendation (SISR) baselines. 
For SR baselines, 
\textbf{SASRec}~\cite{KangM18SASRec} adopts the uni-directional self-attention method to capture the user interest. 
\textbf{DuoRec}~\cite{QiuHYW22DuoRec} enhances SASRec~\cite{KangM18SASRec} with contrastive learning.
\textbf{FMLPRec}~\cite{ZhouYZW22FMLPRec} proposes a filter-enhanced MLP to eliminate frequency domain noise.
\textbf{BSARec}~\cite{aaai/Shin0WP24BSARec} leverages the Fourier transform to inject an inductive bias for modeling user patterns.
For SISR baselines, \textbf{GRU4Rec$_{\text{F}}$} and \textbf{SASRec$_{\text{F}}$} are enhanced versions of GRU4Rec~\cite{HidasiKBT15GRU4Rec} and SASRec~\cite{KangM18SASRec}. Following the previous work~\cite{Lin24MSSR}, the item ID and attributes are fused before feeding to the model via summation and concatenation for GRU4Rec$_{\text{F}}$ and SASRec$_{\text{F}}$, respectively. 
\textbf{S$^3$-Rec}~\cite{ZhouWZZWZWW20S3Rec} utilizes mutual information maximization to capture the correlations between items, sequences, and attributes.
\textbf{FDSA}~\cite{ZhangZLSXWLZ19FDSA} adopts late fusion by utilizing multiple self-attention blocks.
\textbf{NOVA}~\cite{Chang21NOVA} adopts non-invasive self-attention mechanism for effective attention learning.
\textbf{DIF-SR}~\cite{XieZK22DIFSR} decouples the attention calculation of item ID and attributes. 
\textbf{DLFSRec}~\cite{LiuD0P023DLFSRec} proposes distribution-based learnable filters to effectively utilize side-information.
\textbf{MSSR}~\cite{Lin24MSSR} models the multiple user representations via a multi-sequence integrated attention layer.
\textbf{ASIF}~\cite{www/Wang24ASIF} utilizes side-information without noisy interference via fused attention with untied position information.

\vspace{1mm}
\noindent
\textbf{Implementation Details}. 
We implement all models on the open-source recommendation framework Recbole~\cite{recbole}~\footnote{\url{https://github.com/RUCAIBox/RecBole}} or published code. All models are optimized using Adam optimizer~\cite{KingmaB14Adam}, and tune the learning rate in $\{10^{-4},10^{-3}\}$. We set the maximum sequence length to 50, and we stop the training if the validation N@20 decreases for ten consecutive epochs. We tune all the hyperparameters on the validation data and report the performance on the test set using the models that show the highest performance on the validation set. 
For the proposed method, we set both the embedding size and batch size to 256, and both the number of layers and heads are set to 2. We set the frequency component split parameter $c$ to 3 for Beauty, Sports, Yelp datasets and 5 for Toys dataset. We also tune the aggregating hyperparameters $\alpha$ among $\{0.1, 0.3, 0.5, 0.7, 0.9\}$ and loss balancing hyperparameter $\lambda$ among $\{1, 5, 10, 20, 50, 100\}$. The fusion function $\text{Fusion}(\cdot)$ is set to gating for the Yelp dataset and concatenation for the Beauty, Sports, and Toys datasets. For the baseline models, we follow the original papers' settings for other hyperparameters of baselines, and we thoroughly tune them if not available.
All results are averaged over five runs with different seeds, and we conducted the significance test using a paired t-test. Our code is available at \url{https://github.com/HyeYoung1218/DIFF}.

\section{Experimental Results}\label{sec:results}

\subsection{Overall Performance}
Table~\ref{tab:overall_performance} reports the performance comparison between DIFF and other baselines in four real-world datasets. The key observations are as follows. 
(i) DIFF consistently achieves state-of-the-art performance on all datasets against the best competitive baseline, improving R@20 and N@20 by up to 14.1\% and 12.5\%, respectively. Especially, DIFF exhibits the best performance against the best competitive SISR baselines, \eg, MSSR~\cite{Lin24MSSR} and ASIF~\cite{www/Wang24ASIF}, yielding average gains of 4.7\% and 11.2\% on R@20 and N@20. This indicates that DIFF successfully avoids noisy patterns and leverages side-information. 
(ii) When compared to SR baselines that do not use side-information, SISR baselines, especially MSSR~\cite{Lin24MSSR}, ASIF~\cite{www/Wang24ASIF}, and DIFF, generally achieve superior performance. It implies that modeling user preferences with rich item context is critical for recommendation performance.
(iii) Although DLFSRec~\cite{LiuD0P023DLFSRec} leverages frequency-based learnable filters for SISR, it does not primarily focus on fusion methods, which results in comparatively lower performance than late and intermediate fusion approaches (\ie, FDSA~\cite{ZhangZLSXWLZ19FDSA}, NOVA~\cite{Chang21NOVA}, DIF-SR~\cite{XieZK22DIFSR}, MSSR~\cite{Lin24MSSR}, and ASIF~\cite{www/Wang24ASIF}). This highlights the importance of a well-designed fusion strategy in achieving superior performance.
(iv) Among SISR baselines, intermediate fusion approaches (\ie, NOVA~\cite{Chang21NOVA}, DIF-SR~\cite{XieZK22DIFSR}, MSSR~\cite{Lin24MSSR}, and ASIF~\cite{www/Wang24ASIF}) generally demonstrate higher performance than early fusion methods (\ie, GRU4Rec$_{\text{F}}$ and SASRec$_{\text{F}}$) and late fusion method (\ie, FDSA~\cite{ZhangZLSXWLZ19FDSA}). Notably, two early fusion methods (GRU4Rec$_{\text{F}}$ and SASRec$_{\text{F}}$) lose performance of up to 23.6\% and 35.0\% performance at N@20 compared to GRU4Rec and SASRec, respectively. This underscores the importance of delicately designed fusion methods when utilizing side-information. 
(v) Among SISR baselines, ASIF~\cite{www/Wang24ASIF} and DIFF demonstrate particularly promising performance compared to other SISR models. This highlights that, in addition to designing an effective fusion method, eliminating noisy correlations between IDs and attributes further enhances performance by ensuring meaningful interactions are captured.

\begin{table}[]\small
\setlength{\tabcolsep}{5.4pt}
\renewcommand{\arraystretch}{0.95}
\caption{Ablation study of DIFF. FNF refers to the Frequency-based Noise Filtering. IF and AF represent ID-centric Fusion and Attribute-enriched Fusion, respectively. Lastly, RA denotes Representation Alignment.} \label{tab:ablation}
\vspace{-3mm}
\begin{tabular}{c|c|cccc|c}
\toprule
 & Metric & w/o FNF & w/o IF & w/o AF & w/o RA & DIFF \\ \midrule
 \multirow{2}{*}{Yelp} & R@20 & 0.1045 & 0.1174 & 0.1185 & 0.1114 & 0.1200 \\
 & N@20 & 0.0512 & 0.0560 & 0.0564 & 0.0542 & 0.0567 \\ \midrule
\multirow{2}{*}{Beauty} & R@20 & 0.1289 & 0.1290 & 0.1334 & 0.1300 & 0.1347 \\
 & N@20 & 0.0615 & 0.0629 & 0.0576 & 0.0585 & 0.0632 \\ \midrule
\multirow{2}{*}{Sports} & R@20 & 0.0843 & 0.0795 & 0.0851 & 0.0827 & 0.0853 \\
 & N@20 & 0.0322 & 0.0375 & 0.0337 & 0.0330 & 0.0381 \\ \midrule
\multirow{2}{*}{Toys} & R@20 & 0.1373 & 0.1357 & 0.1459 & 0.1357 & 0.1420 \\
 & N@20 & 0.0615 & 0.0651 & 0.0598 & 0.0600 & 0.0657 \\ \bottomrule

\end{tabular}
\vspace{-3.5mm}
\end{table}
\begin{figure}[t]\small
\centering
\includegraphics[width=0.78\linewidth]{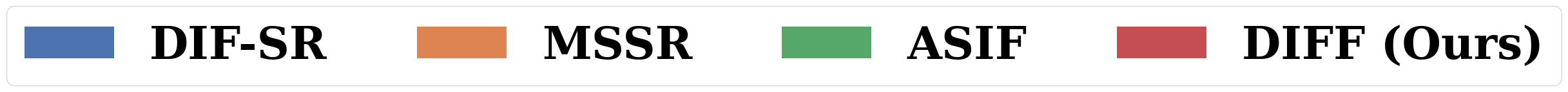}
\begin{tabular}{cc}
\includegraphics[width=0.48\linewidth]{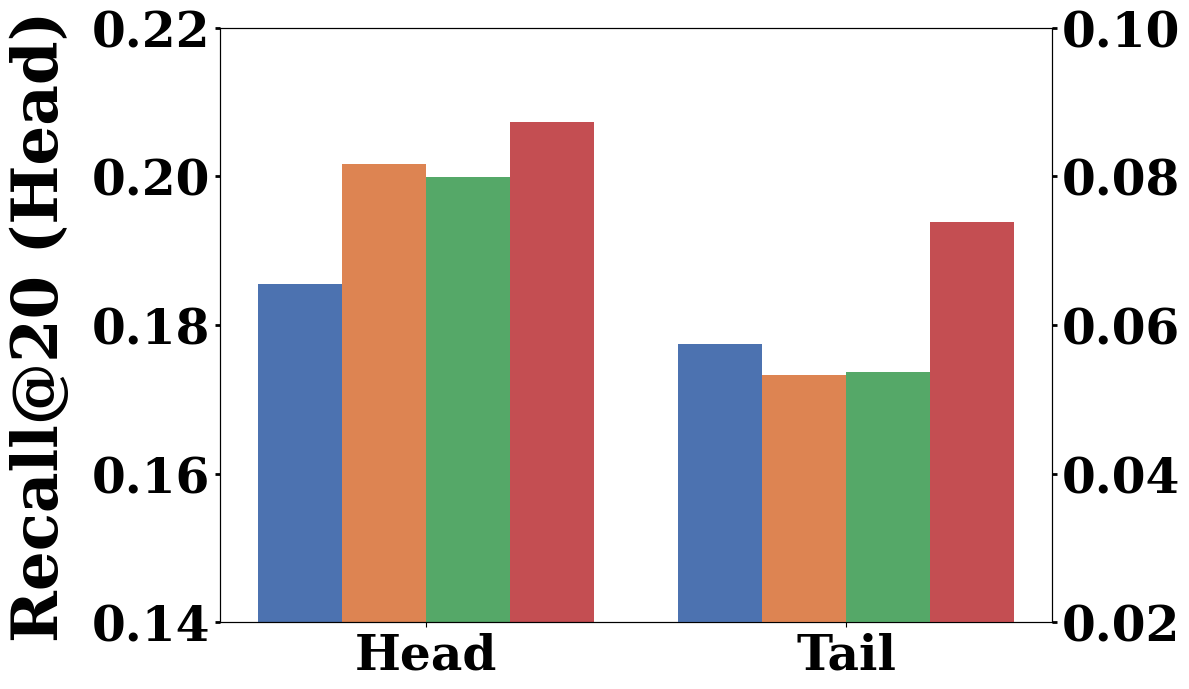} &
\includegraphics[width=0.48\linewidth]{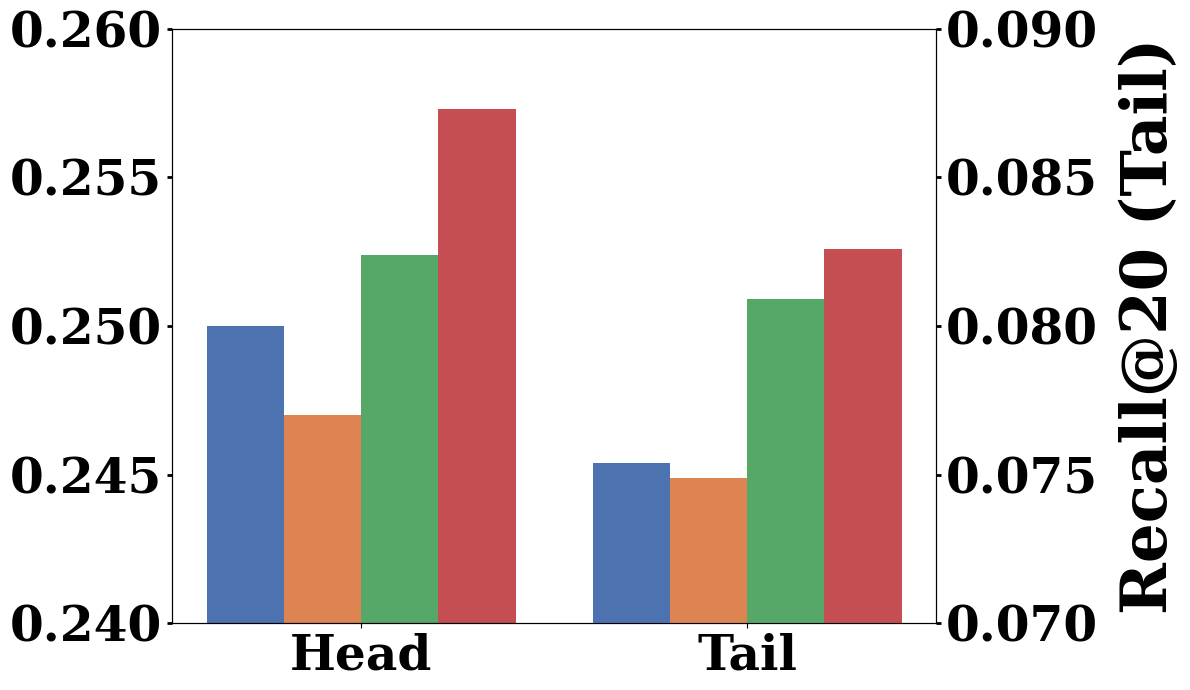} \\
(a) Yelp & (b) Beauty \\

\end{tabular}
\vspace{-4mm}
\caption{Performance comparison on different target item popularity groups. The target items of Head group are the top 10\% most popular items, while the Tail group includes sequences with less popular target items.}\label{fig:exp_headtail}
\vspace{-3mm}
\end{figure}
\begin{figure}[t]\small
\centering
\includegraphics[width=0.78\linewidth]{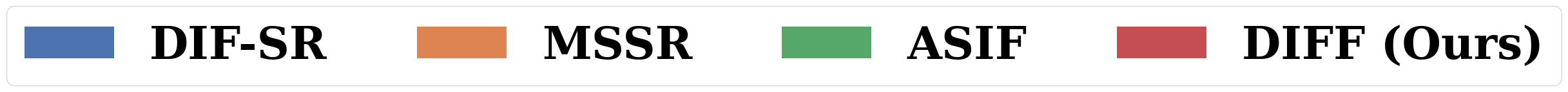}
\begin{tabular}{cc}
\includegraphics[width=0.48\linewidth]{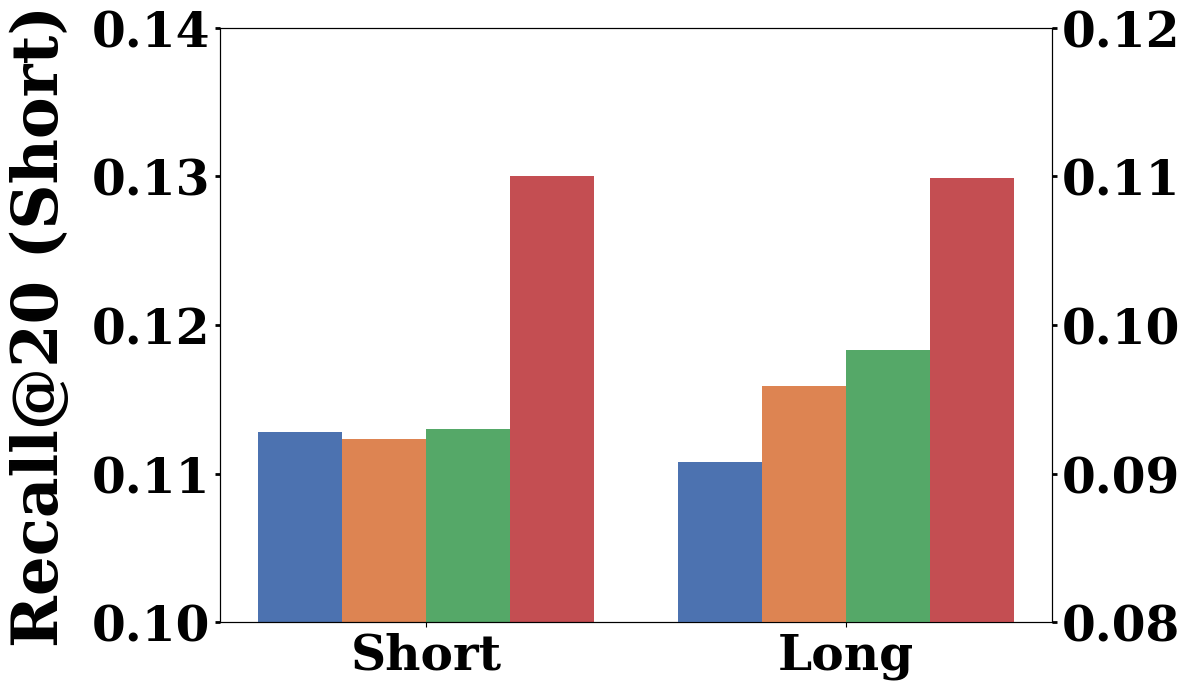} &
\includegraphics[width=0.48\linewidth]{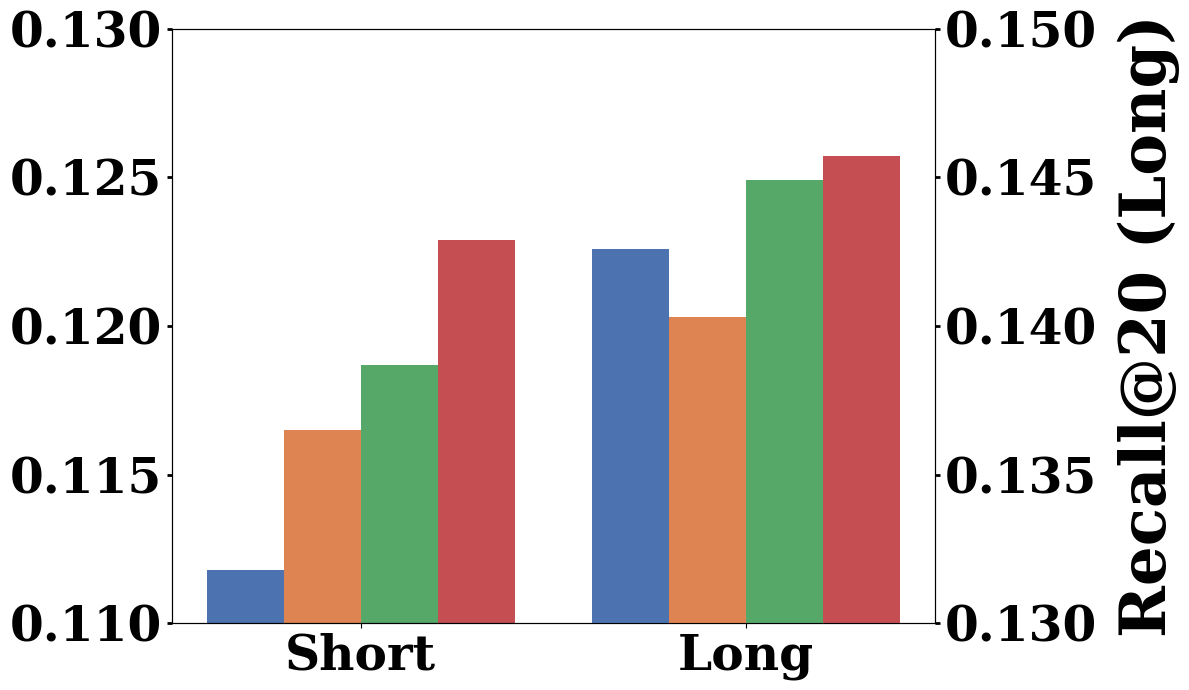} \\
(a) Yelp & (b) Beauty \\
\end{tabular}
\vspace{-4mm}
\caption{Performance comparison on different sequence length groups. The Short group consists of sequences with a length of five (43\% of Yelp and 51\% of Beauty dataset), while the Long group includes sequences longer than five.}\label{fig:exp_seqlen}
\vspace{-3mm}
\end{figure}

\subsection{In-depth Analysis}
\noindent
\textbf{Ablation Study}.
We validate the effectiveness of the key components of the proposed method through the ablation study as shown in Table~\ref{tab:ablation}. 
(i) Frequency-based Noise Filtering (FNF) significantly impacts performance across all datasets, delivering a performance gain of up to 14.8\% and 10.7\% in R@20 and N@20, respectively. It demonstrates that noisy signals are removed and only essential information is successfully extracted, leading to more accurate user representation. 
(ii) The proposed dual fusion strategy remarkably improves the accuracy compared to using only ID-centric Fusion (IF) or Attribute-enriched Fusion (AF) by more than 7.3\% and 13.1\% on R@20 and N@20. It shows that each fusion successfully captures complementary information to another.
(iii) The representation alignment loss (RA) seamlessly integrates item ID and attribute information by aligning their embedding spaces, showing gains of up to 7.7\% and 14.4\% on R@20 and N@20. By harmonizing the representation spaces, ID and attribute information are effectively incorporated into the model.

\noindent
% Effect of Item Popularity
\textbf{Performance by Item Popularity}.
In Figure~\ref{fig:exp_headtail}, we evaluate the performance of DIFF and baseline models by dividing the test user sequences into two groups: Head, consisting of sequences with target items from the top 10\% most popular items, and Tail, with sequences containing less popular target items. The experimental results demonstrate strong performance of DIFF across both groups, effectively alleviating the popularity bias. By leveraging side-information filtering and fusion mechanisms, DIFF can extract meaningful signals from side-information, compensating for the sparse interactions typically associated with tail items. This suggests that DIFF also outperforms other competitive models for cold-start scenarios. In particular, for the Tail group, DIFF achieves up to 38.9\% improvement on Yelp and 10.3\% on the Beauty dataset.

\noindent
% Effect of Sequence Length
\textbf{Performance by Sequence Length}.
In Figure~\ref{fig:exp_seqlen}, we evaluate the performance of DIFF and baseline models by dividing user sequences into two groups based on their length. The Short group consists of users with five interacted items, while the Long group includes users with more than five interacted items. The results show that DIFF consistently outperforms the baseline models across both groups, effectively capturing user preferences regardless of sequence length. Notably, DIFF achieves significant improvements in the Short sequence group, where limited user interaction data. In particular, DIFF shows the performance gains in Recall@20 by up to 15.8\% and 10\% on the Yelp and Beauty dataset, respectively. This indicates that DIFF is particularly effective in scenarios with sparse historical interaction, showcasing its capability to leverage available information more efficiently.

\begin{figure}[t]\small
\centering
\includegraphics[width=0.78\linewidth]{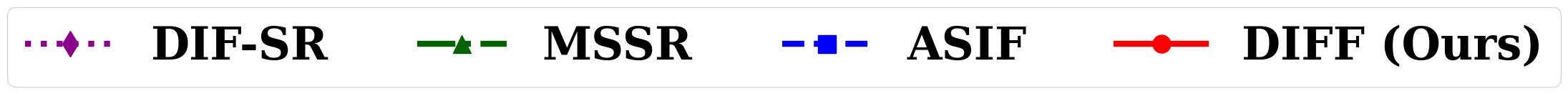}
\begin{tabular}{cc}
\includegraphics[width=0.475\linewidth]{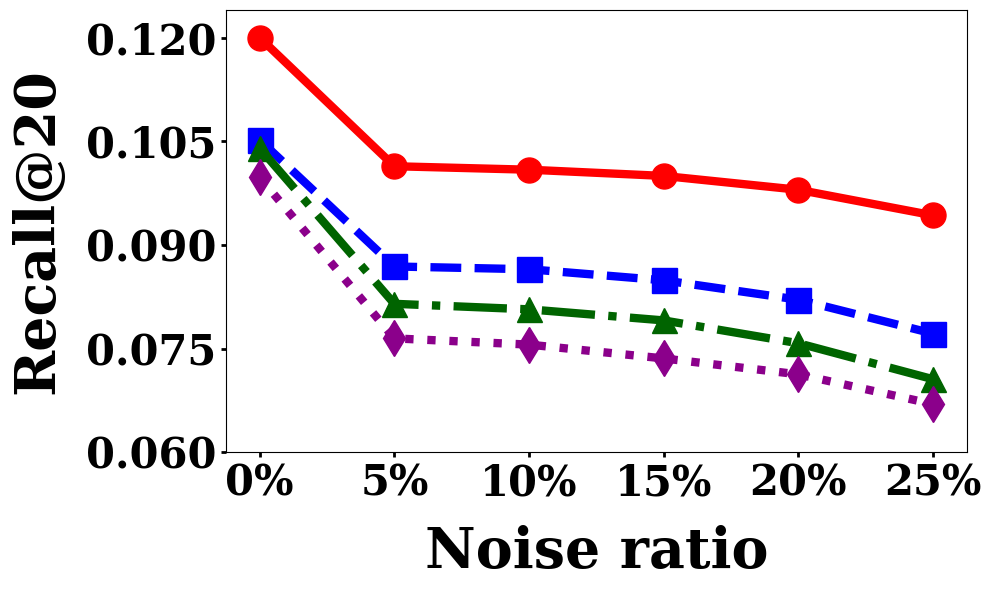} &
\includegraphics[width=0.475\linewidth]{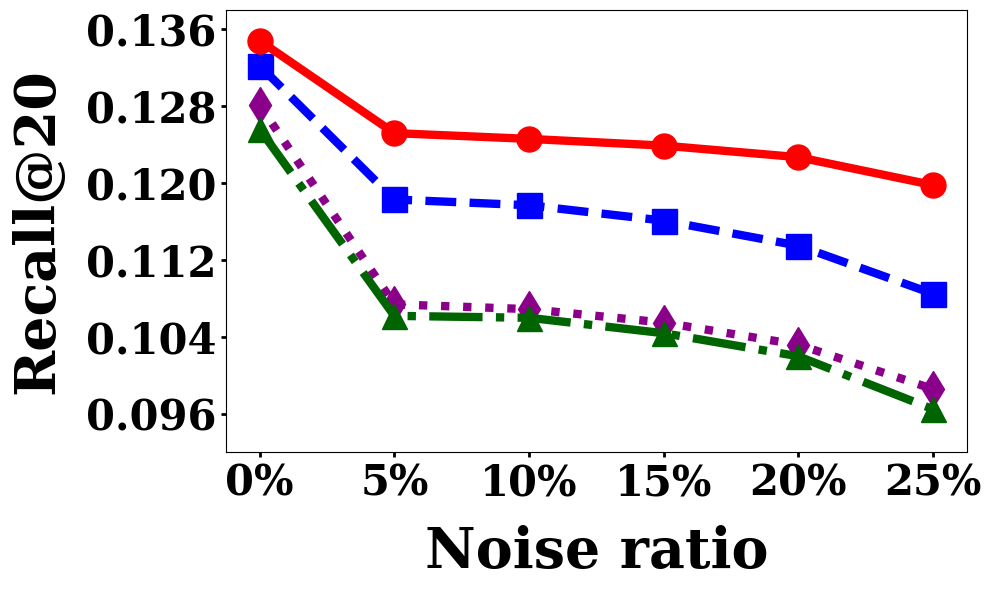} \\
(a) Yelp & (b) Beauty \\
\end{tabular}
\vspace{-4mm}
\caption{Robustness to noisy sequences on Yelp and Beauty datasets. It shows the performance of DIF-SR, MSSR, ASIF, and DIFF by varying the item substitution ratio. }\label{fig:robustness}  
\vspace{-4mm}
\end{figure}

\subsection{Robustness to Noisy Sequence}
In Figure~\ref{fig:robustness}, we examine the robustness of the proposed method to demonstrate the effectiveness of Frequency-based Noise Filtering. Here, we adopt the most competitive SISR models, \ie, DIF-SR~\cite{XieZK22DIFSR}, MSSR~\cite{Lin24MSSR}, and ASIF~\cite{www/Wang24ASIF}, for comparison. Following the approach in ~\cite{DuYZF0LS023SLIME4Rec}, we simulate noisy conditions by injecting synthetic noise into the test sequences. While they add random uniform noise to the original representations, we adopt a more challenging approach by replacing some items in each item ID sequence with random items, resulting in a more realistic and complex evaluation scenario. These substituted items can be regarded as fluctuating items that should ideally be ignored.

The key findings are as follows. (i) Even with a low noise ratio (\ie, 5\%), all models exhibit performance degradation across all datasets, highlighting the challenges posed by noisy inputs. However, DIFF demonstrates greater resilience than DIF-SR, MSSR, and ASIF. Notably, on the Beauty dataset, DIFF shows only a 7.1\% performance drop, whereas ASIF, MSSR, and DIF-SR suffer significant performance drops of 16.2\%, 15.4\%, and 10.5\%, respectively. (ii) As the noise ratio increases incrementally up to 25\%, the performance gap between DIFF and other models consistently widens across all datasets. Notably, on the Yelp dataset, DIFF exhibits a relatively modest decline of 21.4\%, whereas the baseline models show substantial drops, ranging from 26.6\% to 32.9\%. This suggests that baseline models struggle to capture user preferences under noisy conditions. In contrast, our approach effectively filters out noisy signals, ensuring the preservation of critical information.

\subsection{Hyperparameter Sensitivity}\label{sec:results_hyper}
\begin{figure}[t]\small
\centering
\begin{tabular}{cc}
\includegraphics[width=0.48\linewidth]{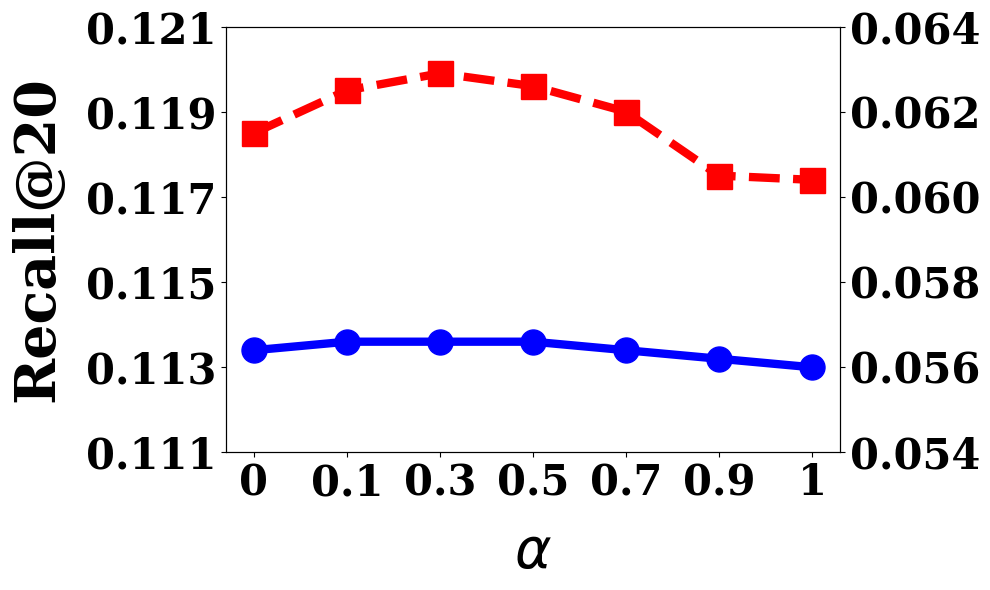} &
\includegraphics[width=0.48\linewidth]{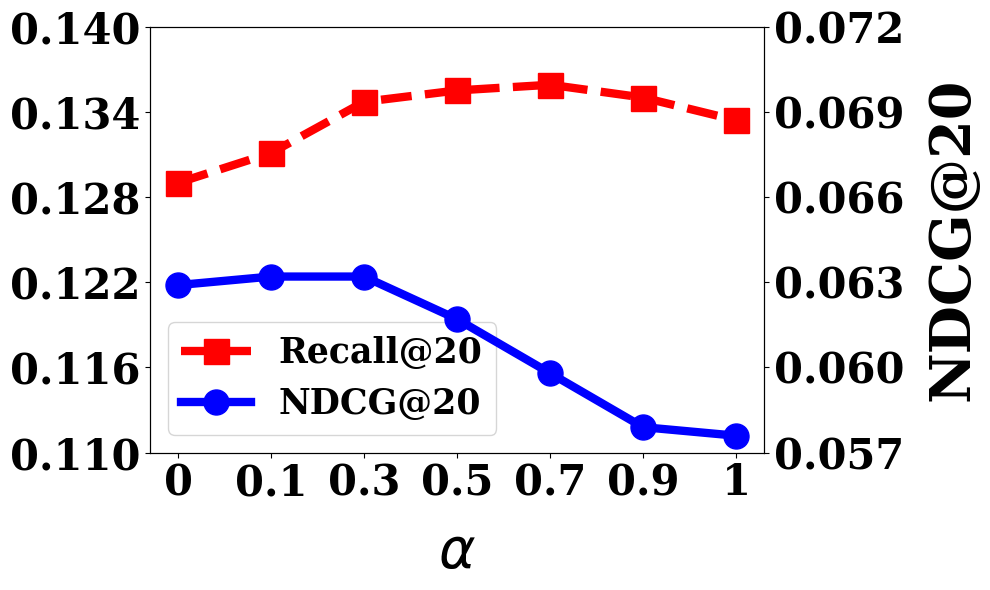} \\
(a) Yelp & (b) Beauty \\
% \includegraphics[width=0.48\linewidth]{Figures/fig_alpha_sports.png} &
% \includegraphics[width=0.48\linewidth]{Figures/fig_alpha_toys.png} \\
% (c) Sports & (d) Toys \\
\end{tabular}
\vspace{-4mm}
\caption{Performance with varying representation aggregating hyperparameter $\alpha$. When $\alpha = 0$, only AF is utilized, and when $\alpha = 1$, only IF is employed.}\label{fig:exp_alpha}
\vspace{-2mm}
\end{figure}

\noindent
\textbf{Representation Aggregating Hyperparameter}.
% alpha=0 -> early (AF) / alpha=1 -> intermediate (IF)
Figure~\ref{fig:exp_alpha} illustrates the sensitivity of the proposed method to representation aggregating hyperparameter $\alpha$ on Yelp and Beauty datasets. In the Yelp dataset, we observe that both Recall and NDCG peak at moderate $\alpha$ values around 0.5, with particularly small fluctuations in NDCG. However, the optimal value of $\alpha$ for the Recall@20 and NDCG@20 performance differs on the Beauty dataset. Higher $\alpha$ values improve recall performance, while lower $\alpha$ values lead to more significant NDCG gains, indicating complementary roles of two fusion types. Our dual fusion approach can effectively enhance performance by leveraging the distinct fusion characteristics.

\begin{figure}[t]\small
\centering
\begin{tabular}{cc}
\includegraphics[width=0.48\linewidth]{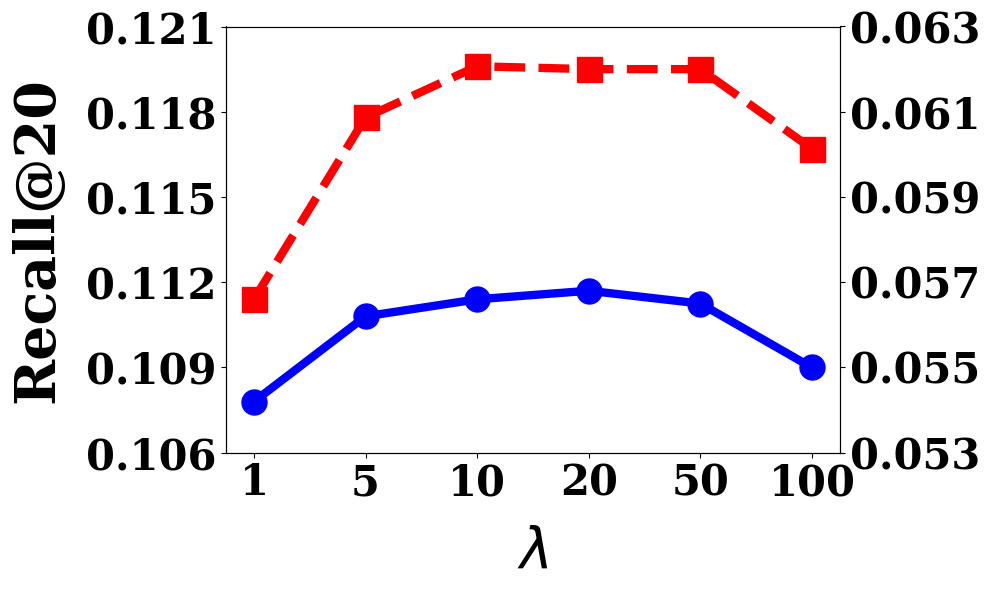} &
\includegraphics[width=0.48\linewidth]{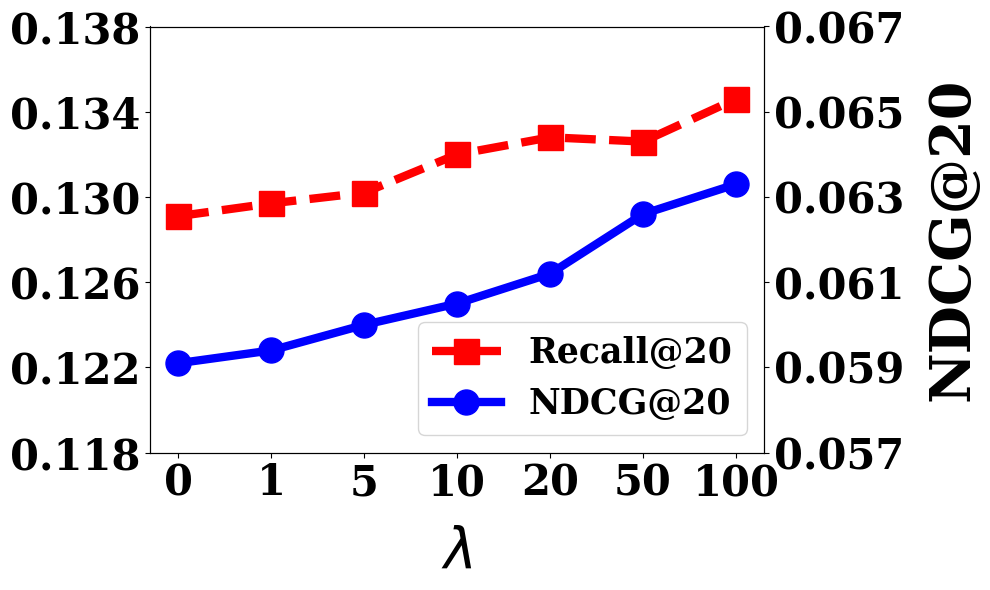} \\
(a) Yelp & (b) Beauty \\

\end{tabular}
\vspace{-4mm}
\caption{Performance with varying alignment loss balancing hyperparameter $\lambda$.}\label{fig:exp_lambda}
\vspace{-2.5mm}
\end{figure}

\noindent
\textbf{Loss Balancing Hyperparameter}.
Figure~\ref{fig:exp_lambda} presents the impact of the loss balancing hyperparameter $\lambda$ across four datasets. The results demonstrate that incorporating the alignment loss consistently improves performance across all datasets. Specifically, we observe performance gains of up to 4.6\% and 6.9\% in NDCG@20 on the Yelp and Beauty datasets, respectively. The Yelp dataset shows a steady increase and peak performance at $\lambda=20$, after which performance decreases slightly. This indicates that an excessively high $\lambda$ may result in over-aligning, which has less impact on performance improvement. For the Beauty dataset, increasing $\lambda$ results in consistent improvements in Recall@20 and NDCG@20, suggesting that greater alignment contributes to better fusion of diverse features. These findings suggest that an optimal $\lambda$ value is crucial for balancing alignment and performance, with different datasets exhibiting varying sensitivities to this hyperparameter.

\subsection{Case Study}
\begin{figure}
\centering

\includegraphics[width=0.96\linewidth]{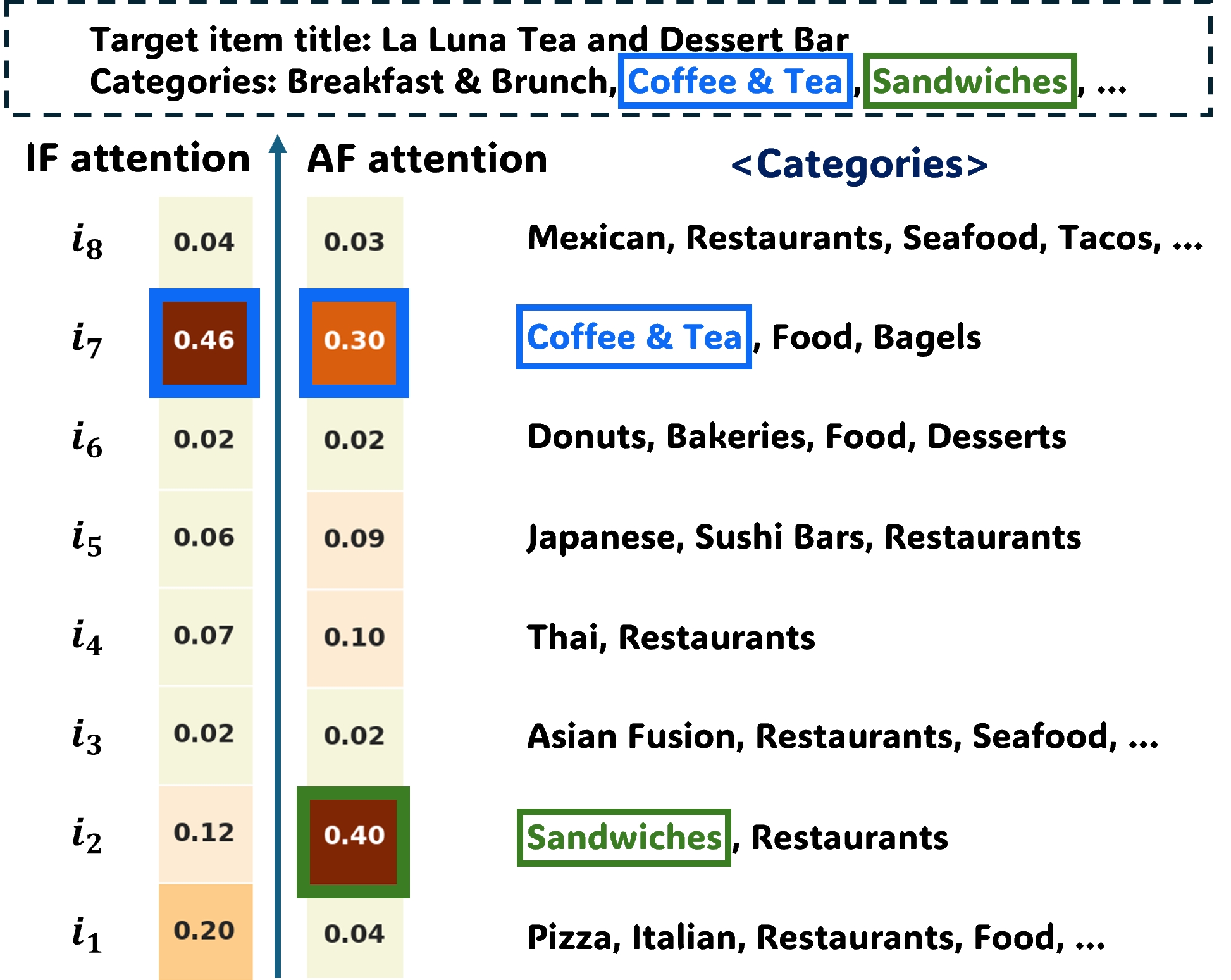}
\vspace{-2mm}
\caption{Case study of attention distribution in the dual fusion types of DIFF, \ie, ID-centric fusion (Left) and Attribute-enriched fusion (Right), on the Yelp dataset.}\label{fig:case_study}
\vspace{-2.5mm}
\end{figure}

In Figure~\ref{fig:case_study}, we conducted a case study on the Yelp dataset to analyze the effectiveness of dual fusion strategies in capturing user preferences. We explore the distribution of attention weights from two fusion strategies, \ie, ID-centric Fusion (IF) and Attribute-enriched Fusion (AF) to understand their individual contributions to the recommendation process. (i) IF allocates the highest attention weight to $i_{7}$ sharing a category of ``\textit{Coffee \& Tea}'' with the target item, demonstrating the ability to prioritize relevant attributes. However, IF alone fails to capture $i_{2}$, which shares a different but relevant category with the target item. (ii) AF allocates high attention weight to $i_{2}$ and $i_{7}$, which shares the ``\textit{Sandwiches}'' category with the target item, covering more diverse items. However, AF alone does not emphasize $i_{7}$ as strongly as IF does, potentially overlooking highly relevant items. These observations indicate that the two fusion strategies capture different aspects of user preferences, with IF excelling at reinforcing specific attribute relevance and AF offering a more diverse coverage. Therefore, the complementary strengths of AF and IF suggest a synergistic potential in combining them into a dual fusion.

\section{Conclusion}\label{sec:conclusion}

In this paper, we introduce the novel \textit{\textbf{D}ual Side-\textbf{I}nformation \textbf{F}iltering and \textbf{F}usion model (\textbf{DIFF})} model, which aims to effectively eliminate noisy interference and fully leverage side-information. For that, DIFF consists of a two-fold process: \textit{Frequency-based Noise Filtering} and \textit{Dual Multi-sequence Fusion}. It is essential to filter inconsistent patterns when incorporating various side-information, ensuring that only the most relevant signals contribute to learning user preferences. Additionally, we successfully combine intermediate and early fusion by leveraging ID-centric and attribute-enriched interactions. Our empirical evaluation reveals that DIFF achieves new state-of-the-art performance by up to 14.1\% and 12.5\% gains in Recall@20 and NDCG@20 across four benchmark datasets.

\begin{acks}
    This work was partly supported by the Institute of Information \& communications Technology Planning \& evaluation (IITP) grant and the National Research Foundation of Korea (NRF) grant funded by the Korea government (MSIT) (No. RS-2019-II190421, RS-2021-II212068, RS-2022-II220680, RS-2025-00564083, and IITP-2025-RS-2024-00437633, each contributing 20\% to this research).
\end{acks}

\newpage
\bibliographystyle{ACM-Reference-Format}
\balance
\bibliography{references}

\end{document}